\documentclass[aps,reprint,superscriptaddress,amsmath,amssymb,prb,]{revtex4-2}
\usepackage{amsmath} 
\usepackage{xcolor}
\usepackage[colorlinks=true]{hyperref}
\hypersetup{linkcolor = [RGB]{206,73,107},
            citecolor = [RGB]{43, 105, 178},
            urlcolor = [RGB]{43, 105, 178}}
\usepackage{graphicx}
\usepackage{physics}
\usepackage{dcolumn}
\usepackage{xfrac}
\usepackage{bm}
\usepackage{hyperref}
\usepackage{lipsum}
\usepackage{array, tabularx,  ragged2e,  booktabs}
\usepackage[version=4]{mhchem}
\newcommand{\ov}{OV$^0$}
\newcommand{\us}{$\mathrm{\mu}$s}

\begin{document}
\preprint{APS/123-QED}
\title{A defect in diamond with millisecond-scale spin relaxation time at room temperature}

\author{Sounak Mukherjee}
\author{Anran Li}
\author{Johannes Eberle}
\author{Sean Karg}
\author{Zi-Huai Zhang}
\affiliation{Department of Electrical and Computer Engineering, Princeton University, Princeton, NJ 08544, USA}
\author{Mayer M. Feldman}
\affiliation{Department of Physics, Princeton University, Princeton, NJ 08544, USA}
\author{Yilin Chen}
\affiliation{Materials Department, University of California, Santa Barbara, CA 93106, USA}
\author{Mark E. Turiansky}
\affiliation{Materials Department, University of California, Santa Barbara, CA 93106, USA}
\affiliation{US Naval Research Laboratory, 4555 Overlook Ave SW, Washington, DC 20375, USA}
\author{Mengen Wang}
\affiliation{Materials Department, University of California, Santa Barbara, CA 93106, USA}
\affiliation{Department of Physics and Astronomy, University of North Carolina at Chapel Hill, Chapel Hill, NC 27599, USA}
\author{Yogendra Limbu}
\author{Tharnier O. Puel}
\author{Yueguang Shi}
\affiliation{Department of Physics and Astronomy, University of Iowa, Iowa City, IA 52242, USA}
\author{Matthew L. Markham}
\author{Rajesh L. Patel}
\affiliation{Element Six, Harwell OX11 0QR, United Kingdom}
\author{Patryk Gumann}
\affiliation{IBM T.J. Watson Research Center, Yorktown Heights, NY 10598, USA}
\author{Michael E. Flatt\'e}
\affiliation{Department of Physics and Astronomy, University of Iowa, Iowa City, IA 52242, USA}
\author{Chris G. Van de Walle}
\affiliation{Materials Department, University of California, Santa Barbara, CA 93106, USA}
\author{Stephen A. Lyon}
\author{Nathalie P. de Leon}
\email{npdeleon@princeton.edu}
\affiliation{Department of Electrical and Computer Engineering, Princeton University, Princeton, NJ 08544, USA}

\date{\today}

\begin{abstract}
Spin defects in diamond are promising platforms for quantum sensing. The longest electron spin relaxation times ($T_1$) at room temperature for solid-state defects are observed in nitrogen vacancy centers in diamond, which can reach 6.67~ms \cite{cambria2023temperature}, and substitutional nitrogen (``P1 centers") in diamond, which exhibit a $T_1$ of 2~ms \cite{carroll2021electron}. No other solid-state defect has exhibited millisecond-scale spin relaxation times at room temperature thus far. Here, we characterize the spin properties of the WAR5 defect in diamond \cite{cann2009magnetic} with pulsed electron spin resonance. The observed $T_1$ is one of the longest for solid-state spin defects: 0.97(27)~ms at room temperature and 14.38(19)~min at 4~K. The observed coherence time ($T_2$) is 246(7)~\us, which can be extended to 6.49(34)~ms at 4~K with dynamical decoupling. Furthermore, we demonstrate optical spin polarization with a range of wavelengths from 405~nm to 500~nm and propose potential zero-phonon line candidates. 

\end{abstract}
\maketitle
\section{Introduction}
Spin defects in solid-state hosts are widely explored for a variety of quantum applications, including quantum sensing and quantum information processing. For quantum sensing, achieving a high sensitivity requires long spin coherence ($T_2$) of the defect, which is fundamentally limited by the spin-lattice relaxation time ($T_1$) \cite{wolfowicz2021quantum}. Additionally, optically addressability at room temperature and ambient conditions opens up the platform for a wide range of sensing modalities across different materials. The negatively charged nitrogen vacancy (NV) center in diamond is the most successful among such defects by virtue of its extraordinary spin properties, enabling sensing applications in magnetometry \cite{taylor2008high, rovny2024nanoscale}, electric field sensing \cite{dolde2011electric}, and thermometry \cite{kucsko2013nanometre}. It exhibits the longest $T_1$ of 6.67 ms \cite{cambria2023temperature} among all electron spin defects in solid state at room temperature.

This extraordinary room-temperature spin lifetime motivates the question of how unique NV centers in diamond are, and whether other optically-addressable solid-state defects exist that have millisecond-scale spin lifetime in ambient conditions. The discovery of such defects could enable new technologies, such as novel sensing modalities and multimodal sensing. Diamond is an exceptional host material with the highest Debye temperature and low spin-orbit coupling, which allows long spin relaxation times at elevated temperatures \cite{wolfowicz2021quantum}. Combined with high-purity synthesis and isotopic purification, a long $T_1$ enables long spin coherence times. In recent years, new defects like the group-IV vacancy centers in diamond have emerged, including the negatively charged silicon, tin, and germanium vacancies \cite{rogers2014all, becker2018all, trusheim2020transform, siyushev2017optical}, and the neutral silicon vacancy center \cite{green2017neutral, rose2018observation}, none of which reach millisecond-scale $T_1$ at room temperature. Substitutional nitrogen (P1) centers have long $T_1$ at room temperature ($\sim$2~ms) \cite{carroll2021electron, reynhardt1998temperature}, but they lack optical spin polarization and readout. Among defects in other materials, the longest $T_1$ at room temperature has been reported for silicon vacancies (V2) in silicon carbide (SiC) with 500~$\mu$s \cite{widmann2015coherent,simin2017locking}, followed by PL6, PL8, and PL5 defects in SiC with 230~$\mu$s, 188~$\mu$s,  and 158~$\mu$s, respectively \cite{yan2020room, hu2024room, li2022room}, and 18~$\mu$s for boron vacancy (V$_\text{B}^-$) in hexagonal boron nitride (hBN) \cite{gottscholl2020initialization}. Furthermore, some alternative defects have been proposed from theoretical calculations to have NV-like properties \cite{davidsson2025nv, groppfeldt2025high, abbas2025theoretical, harris2020group, turiansky2023telecom, umeda2022negatively}.

A natural strategy for searching for solid-state defects with long $T_1$ at room temperature is to explore centers with a similar structure and composition to NV centers. The WAR5 defect, first reported in Ref.~\citealp{cann2009magnetic}, is hypothesized to be the neutral oxygen vacancy (\ov{}) center \cite{thiering2016characterization}, which is isolectronic to NV$^-$. Hence, \ov{} is expected to share a similar electronic structure and spin properties with the NV center, which we explore in this work. Density functional theory \cite{thiering2016characterization, zhang2014neutral} predicts \ov{} to have $C_{3v}$ symmetry, $S=1$ ground state, and a zero-field splitting (ZFS) of $D=2.989$ GHz. The magnetic resonance spectrum and ground state spin Hamiltonian were reported in Ref.~\citealp{cann2009magnetic}, but thus far there has been no characterization of the spin dynamics or optical spin polarization of WAR5.

In this paper, we report that WAR5 exhibits one of the longest $T_1$s measured for an electronic spin in a solid-state host at room temperature. First, we obtain the g-factor and the ZFS of WAR5 from the orientation dependence of electron spin resonance (ESR) spectra at cryogenic temperatures. We characterize the spin dynamics of WAR5 with pulsed ESR, and show that the temperature dependence of $T_1$ is governed by direct and Orbach processes. We also analyze the spin decoherence and deduce that the $T_2$ is limited by the noise bath of P1 centers in the sample. With dynamical decoupling, $T_2$ is extended to milliseconds at 4~K. As a first step towards optical addressability, we demonstrate optical spin polarization (OSP) of the defect and investigate potential zero-phonon line (ZPL) candidates observed in photoluminescence (PL). Finally, we propose a model for the OSP from electronic structure calculations of \ov.

\section{Results and Discussion}
\begin{figure}[t]
\includegraphics[width=1\linewidth]{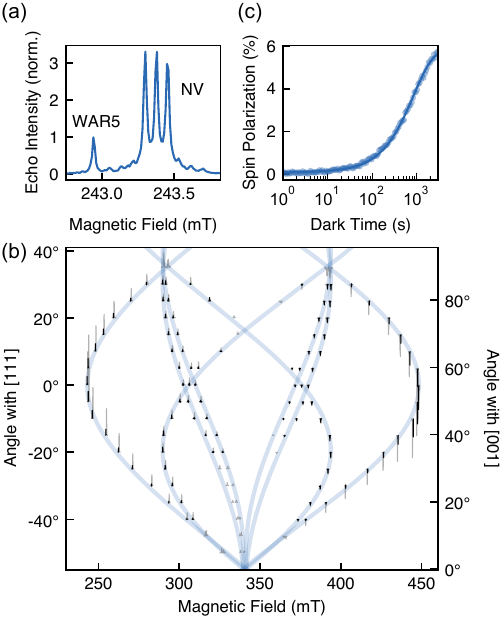}
    \caption{(a) Pulsed ESR spectrum of the WAR5 sample at 4~K with the magnetic field aligned to $\langle111\rangle$. Optical polarization with 455~nm shows the WAR5 peak at 242.95~mT ($m_s=0\leftrightarrow+1$ transition). The set of three peaks centered at 243.38 mT originates from NV centers in the same sample, with their characteristic hyperfine splitting due to $^{14}$N nuclei. (b) WAR5 ESR as a function of the angle between the crystal axes and the magnetic field. ESR spectra including the NV peaks are shown in gray. WAR5 transitions are extracted from multi-Lorentzian fits and plotted in black. Blue lines are fits to the WAR5 Hamiltonian with g-factor and zero-field splitting as free parameters. All transitions are corrected to the same resonance frequency, 9.696~GHz. (c) Saturation recovery on WAR5 at 4~K. Exponential fit gives $T_1$ = 14.38(19)~min.}
    \label{fig1}
\end{figure}

All experiments in this work have been carried out on a diamond grown by chemical vapor deposition, hereby referred to as the WAR5 sample. The sample was synthesized using a high oxygen chemistry along with nitrogen doping at Element Six \cite{cann2009magnetic}. Pulsed ESR measurements in the X-band were performed in a home-built setup with a helium flow cryostat (Appendix \ref{appendix:esr}). Spin polarization is investigated using a multimode fiber in contact with the sample, which delivers 10 to 70~mW optical power from fiber-coupled lasers chosen for the wavelength of interest.

\subsection{Ground state Hamiltonian characterization}
Fig.~\ref{fig1}(a) shows the pulsed ESR spectrum of the WAR5 sample at 4~K with the magnetic field aligned to a $\langle111\rangle$ axis of the diamond. A 455~nm laser is used to optically polarize the spins for 10~s. The higher intensity peaks centered at 243.38~mT belong to the NV center, with its characteristic set of three peaks arising from hyperfine interactions with $^{14}$N nuclei \cite{felton2009hyperfine}. The peak at 242.95~mT originates from a different spin species with $S=1$, which exhibits a different optical spin polarization spectral response from NV [Fig.~\ref{fig:esr_comparison}(a)]. This defect is identified as WAR5, which was previously observed in continuous wave ESR \cite{cann2009magnetic}. The growth chemistry of the sample suggests that this is an oxygen-related defect, consistent with the lack of hyperfine splitting, as the natural abundance of oxygen is predominantly  $^{16}$O with no nuclear spin. Fig.~\ref{fig1}(b) shows WAR5 and NV transitions as a function of the angle between crystal axes and magnetic field, when the sample is rotated about a $\langle110\rangle$ axis. The WAR5 defect has the same form of the ground-state spin Hamiltonian as the NV center \cite{felton2009hyperfine},
\begin{equation}
H = \mu_B \mathbf{B} \mathbf{\hat{g}} \mathbf{S} + \mathbf{S}\hat{\mathbf{D}} \mathbf{S}
\end{equation}
where $\mu_B$ is the Bohr magneton, $\mathbf{B}$ is the magnetic field, $\mathbf{S}$ is the electron spin operator vector, $\hat{\mathbf{D}}$ is the ZFS tensor, and $\mathbf{\hat{g}}$ is the electron g-tensor. Each of the four ZFS axes along $\langle111\rangle$ gives rise to two transitions ($m_s=0\leftrightarrow\pm1$), leading to eight lines for WAR5 at a given crystal orientation. Imperfect sample mounting causes the rotation axis to be slightly tilted from $\langle110\rangle$ (Appendix \ref{appendix:esr}). We extract a misalignment of $3.1(1)^\circ$ and a magnetic field offset of 0.16(9)~mT by fitting the NV transitions with known ZFS \cite{cambria2023physically} and g-factor \cite{felton2009hyperfine}. Assuming an axial ZFS and isotropic g-tensor, we fit the WAR5 transitions to find $D=2.8877(52)$~GHz and $g=2.0030(6)$ at 4~K. These parameters are consistent with the predicted ZFS and g-factor of \ov{} \cite{thiering2016characterization}. Further supported by electronic structure calculations, we tentatively assign the WAR5 defect as \ov{}.

\subsection{Spin relaxation and coherence}
\begin{figure}[b]
\includegraphics[width=1\linewidth]{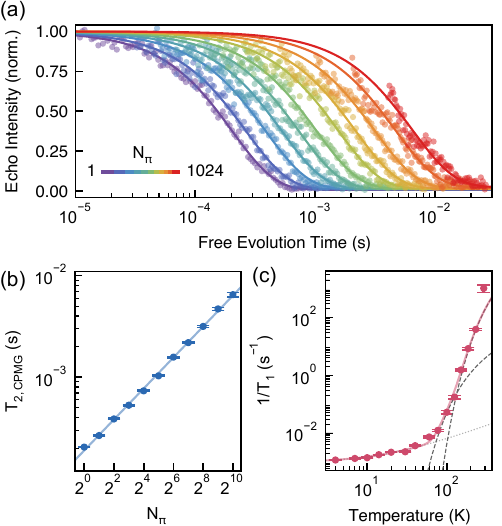} 
    \caption{(a) Coherence of WAR5 under dynamical decoupling using the CPMG sequence with number of $\pi$-pulses, $N_\pi$ = 1, 2, 4, \dots, 1024 (purple to red) at 4~K. For 1024 $\pi$-pulses, $T_{2,\rm{CPMG}}$ = 6.49(34)~ms. (b) $T_{2,\rm{CPMG}}$ as a function of $N_\pi$. The solid line is a fit to the data points yielding a scaling power of 0.507(6), which shows that $T_{2,\rm{CPMG}}$ scales as $\sqrt{N_\pi}$. (c) Temperature dependence of $T_1$ measured using saturation recovery. The solid line is a fit to the data with Eq.~\ref{eq:T1}, excluding the room-temperature measurement. The dotted and dashed gray lines correspond to the direct and Orbach components of the fitted curve, respectively.}
    \label{fig2}
\end{figure}

We now proceed to characterize the spin dynamics of the WAR5 defect with pulsed ESR. Conventionally, the inversion recovery sequence is used to measure $T_1$ in ESR, where spins polarized in the $m_s=0$ state are flipped with a $\pi$-pulse and allowed to relax from $m_s=+1$ to $m_s=0$. However, due to the high density of spins in the WAR5 sample, spectral diffusion of spins \cite{schweiger2001principles} leads to a biexponential decay in inversion recovery (Appendix \ref{appendix:T1}). We mitigate this effect by measuring $T_1$ with saturation recovery instead. In this sequence, a series of $\pi/2$ pulses, also known as a picket fence sequence \cite{eaton2005saturation}, is utilized to saturate the spin polarization after an optical polarization step, and we record the recovery of spin polarization to thermal equilibrium. WAR5 shows long spin relaxation times across a large range of temperatures. The $T_1$ is observed to be 14.38(19)~min at 4~K [Fig.~\ref{fig1}(c)], 1.35(11)~min at 77~K, and 0.97(27)~ms at room temperature [Fig.~\ref{fig:T1spectral}(e)]. We examine the temperature dependence of $T_1$ to determine the mechanisms driving spin relaxation. In Fig.~\ref{fig2}(c), we observe that the relaxation rate increases linearly until $\sim$60~K, followed by an exponential rise at elevated temperatures. Hence, at low temperatures, spin-lattice relaxation of WAR5 can be explained by the direct process via single phonons. At higher temperatures, the $T_1$ is dominated by the Orbach process with two phonon modes. In contrast to NV centers, where a similar behavior originates from quasilocalized phonon modes associated with nitrogen \cite{cambria2023temperature}, we find that in \ov{}, the vibrations of oxygen and carbon atoms contribute to the lower and higher energy phonon groups, respectively (Appendix \ref{appendix:phonon}). Thus, the temperature dependence of the relaxation rate can be modeled by:
\begin{equation}
    \frac{1}{T_1} = \frac{1}{T_{1_\text{sat}}} + \Gamma_\text{direct}T + \frac{\Gamma_{\text{Orbach}_1}}{e^{E_1/k_B T}-1} + \frac{\Gamma_{\text{Orbach}_2}}{e^{E_2/k_B T}-1}
\label{eq:T1}
\end{equation}
Fitting the data in Fig.~\ref{fig2}(c) yields activation energies, $E_1=54.3(8.7)$~meV and $E_2=137.8(10.9)$~meV, and coupling rates, $\Gamma_{\text{Orbach}_1}=28(5)$~Hz, $\Gamma_{\text{Orbach}_2}=38.44(21.60)$~kHz, and $\Gamma_\text{direct} = 5.5(3.1)\times10^{-5}$~Hz/K, consistent with theoretical calculations (Fig.~\ref{fig:pdos}). The $T_1$ levels off near liquid helium temperatures with $T_{1_\text{sat}}=17.51(2.84)$~min. This limiting rate has been shown to be sample-dependent for the case of NV due to cross-relaxation arising from dipole-dipole interaction \cite{jarmola2012temperature}. WAR5 exhibits a higher $T_1$ compared to NV centers in the WAR5 sample at 4~K. As we go to higher temperatures, the relaxation time falls off faster due to the stronger coupling to higher energy phonons ($\Gamma_{\text{Orbach}_2}$), and we observe the WAR5 $T_1$ to be lower than NV centers at room temperature. We note that the relaxation rate at 290~K is slightly higher than the model predicts, which may indicate other contributions to spin relaxation at higher temperatures, such as Raman processes.

\begin{figure*}[httbp]
\includegraphics[width=1\linewidth]{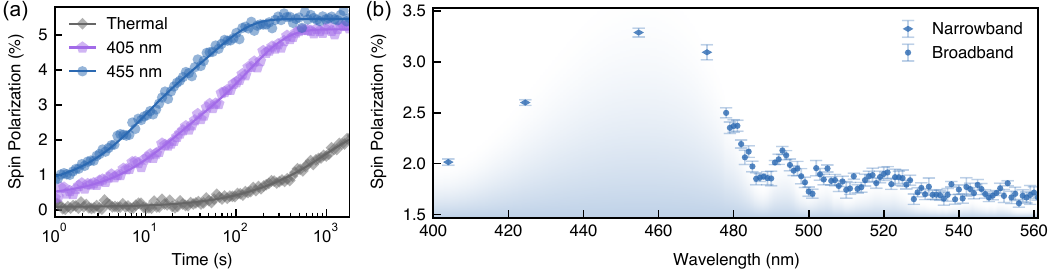} 
    \caption{Optical spin polarization of WAR5 at 10~K. (a) Spin polarization as a function of the polarization time for optical and thermal polarization. 50~mW of optical power is used. The solid curves are biexponential fits, which yield time constants of 55(4)~s, 2.3(1)~min, and 16.3(1.8)~min for 455~nm, 405~nm, and thermal polarization, respectively. The shorter timescale is attributed to spectral diffusion. (b) Optical spin polarization as a function of wavelength with 10~mW of optical power. Spins are polarized for 2~min, which corresponds to a thermal polarization baseline of 0.4\%. Broadband and narrowband measurements correspond to 20~nm and $\lesssim1$~nm spectral bandwidths, respectively. The discontinuity at 501~nm is from switching bandpass filters. Error bars are standard errors from averaging 5 measurements. The shaded region is a guide for the eye to indicate the absorption spectrum of WAR5.}
    \label{fig3}
\end{figure*}

Next, we characterize the coherence times of the defect using the Hahn echo sequence. The $T_2$ for WAR5 is 246(7)~$\mu$s at 4~K and 4.9(7)~$\mu$s at room temperature [Fig.~\ref{fig:T2noise}(d)]. These are much shorter than that expected from the $2T_1$ limit. Fig.~\ref{fig:T2noise}(a) shows the coherence of the defect fitted to a stretched exponential, $e^{-(t/T_2)^{\beta}}$, where $\beta$ is the stretching factor. For WAR5 at 4~K, we obtain $\beta\approx1.22$. Notably, NV centers in the WAR5 sample show a similar $T_2$ of 183(2)~\us, with $\beta\approx1.44$ [Fig.~\ref{fig:T2noise}(b)]. This is much lower than the typical $T_2$ reported for NV centers at cryogenic temperatures \cite{cambria2023temperature}. Hence, we conclude that the same noise source limits both NV and WAR5 coherence. Ensemble spin coherence for NV centers in diamonds with a high density of electron spins has been observed to be spin bath limited, showing a linear scaling with the P1 concentration \cite{bauch2020decoherence, park2022decoherence}. The stretching factor for such an ensemble can be calculated to be 1.5 from a Lorentzian spin bath model \cite{medford2012scaling}. Experimentally, it is measured to be in the range of 1.2–1.5 \cite{marcks2024guiding}, which is consistent with the stretching factor observed for NV in the WAR5 sample. Generalized correlated-cluster expansion (gCCE) methods \cite{ghassemizadeh2024coherence} predict the stretching factor for NV ensemble decoherence to be around 1.2 to 1.3 at high densities of P1 centers. We simulate \ov{} decoherence under the observed noise bath using gCCE (Appendix \ref{appendix:cce}), which yields a stretching factor of 1.23 [Fig.~\ref{fig:T2noise}(c)], in excellent agreement with our experimental result.

To further study the noise bath and extend the coherence times of WAR5, we use dynamical decoupling. $T_2$ measured using the Carr-Purcell-Meiboom-Gill (CPMG) sequence for varying number of $\pi$-pulses, $N_\pi$, is shown in Fig.~\ref{fig2}(a). To access short interpulse spacing times ($\tau$) required for higher order CPMG, we overcouple the resonator and use phase cycling to overcome the dead time due to the resonator ring-down (Appendix \ref{appendix:esr}). The maximum $T_{2,\rm{CPMG}}$ is achieved to be 6.49(34)~ms at 4~K for 1024 $\pi$-pulses, limited only by the minimum $\tau$ which can be used in the setup. $T_{2,\rm{CPMG}}$ exhibits a power law scaling with the number of pulses, $T_{2,\rm{CPMG}} \propto \sqrt{N_\pi}$ [Fig.~\ref{fig2}(b)]. Using CPMG as a filter function, we can compute the noise spectrum $S(\omega)$ of the spin bath \cite{bar2012suppression} [Fig.~\ref{fig:T2noise}(e)]. The P1 spin bath is expected to give rise to a Lorentzian spectrum, governed by Ornstein-Uhlenbeck noise \cite{bar2012suppression, bauch2020decoherence}. In particular, we find that a good fit to the experimental noise spectrum is given by a double Lorentzian model. From the fit parameters and spin counting, we attribute the two Lorentzian noise baths to the P1 electron spins and the natural abundance of $^{13}$C nuclear spins (Appendix \ref{appendix:noise}).

\subsection{Optical spin polarization}
The ability to optically polarize the population to a known spin state is the first step towards achieving optical control of a spin qubit. Spin polarization is defined as $\frac{p_0-p_1}{p_0+p_1}$, where $p_0$ and $p_1$ denote the populations in the $m_s=0$ and $m_s=+1$ states. The ESR signal is normalized to the Boltzmann distribution from thermal polarization with a saturation recovery measurement. We measure the spin polarization as a function of illumination time, as shown in Fig.~\ref{fig3}(a), which depicts how fast the spins can be polarized and the saturated polarization value. The most efficient and fastest OSP is achieved with 455~nm excitation, yielding a maximum of 5.45\%, followed by 405~nm saturating at 5.16\% [Fig.~\ref{fig3}(a)]. Both of these provide substantial gains over thermal polarization, which saturates at 2.33\% at 10~K, with a timescale following the long $T_1$ of WAR5. The curves in Fig.~\ref{fig3}(a) are fit to a biexponential, with the shorter timescale originating from spectral diffusion similar to what is observed in an inversion recovery measurement (Appendix \ref{appendix:T1}). Inversion recovery on $m_s=0\leftrightarrow+1$ vs $m_s=0\leftrightarrow-1$ transitions indicates that wavelengths in the range of 405–500~nm preferentially polarize the spin into the $m_s=0$ state [Fig.~\ref{fig:T1spectral}(d)]. The OSP for WAR5 is found to be maximized when the magnetic field is aligned to a $\langle111\rangle$ crystal axis, similar to NV centers \cite{drake2015influence}. We note that the OSP process for WAR5 is slower than NV centers. With 532~nm, NV saturates to a higher spin polarization (22.5\%) with a characteristic timescale of 1.45~s when measured in the same optical configuration. 

OSP as a function of wavelength is shown in Fig.~\ref{fig3}(b). A broadband supercontinuum laser source is filtered with a high-extinction bandpass filter (20~nm bandwidth), and the central wavelength is scanned by tuning the angle of incidence (Appendix \ref{appendix:sfg}). The supercontinuum laser reaches a lower limit at 477 nm to provide sufficient optical power. Hence, we use separate narrowband laser diodes ($\lesssim$1~nm bandwidth) at fixed wavelengths to measure the OSP at lower wavelengths. Since OSP is related to the absorption spectrum of a defect \cite{zhang2020optically}, we can interpret the broad region from 405~nm to 480~nm as the mirror image of the emission spectrum arising from the phonon sideband of WAR5. Despite the broad spectral bandwidth of this configuration, fine-tuning the wavelength reveals a resonant feature around the ZPL, as demonstrated for NV centers (Fig. \ref{fig:nvosp}). Fig.~\ref{fig3}(b) shows a distinct peak in OSP around 491 nm for WAR5. The broad bandwidth, combined with the inhomogeneity of the excitation spectrum, introduces an uncertainty in wavelength for the resonant feature observed around the ZPL. Thus, from the wavelength dependence of OSP, we deduce that the ZPL is between $\sim$480 and $\sim$500~nm.

\subsection{Zero-phonon line candidates}
Using PL spectroscopy, we observe multiple lines in the range of  450–550~nm in the WAR5 emission spectrum (Fig.~\ref{fig:pl}). The high density of nitrogen and other impurities may result in numerous damage centers, nitrogen-related defects, and unidentified color centers, making it experimentally challenging to isolate the WAR5 ZPL. Detailed information about the observed peaks is available in Appendix \ref{appendix:pl}. 

Scanning a narrow linewidth laser across the ZPL can give rise to a strong response in the ESR signal and confirm the association of a PL center with a spin defect in ESR \cite{green2017neutral, rose2018observation, phenicie2019narrow}. We use this technique to search for the ZPL of WAR5. Since a distinct feature is observed in OSP for WAR5 around 491~nm [Fig.~\ref{fig3}(a)], we primarily investigate the ESR response with narrow linewidth excitation in this region, where peaks at 491~nm and 492~nm are also observable in PL. This is achieved with the help of second harmonic generation (Appendix \ref{appendix:sfg}). However, no resonance is observed within 489–494~nm [Appendix~\ref{appendix:opticalesr}, Fig.~\ref{fig:opticalesr}(a)]. The absence of a resonance feature in the fine wavelength scans does not completely rule out a peak as a ZPL candidate, as sample-specific charge dynamics can complicate resonant OSP. Furthermore, linewidths much narrower than the inhomogeneous distribution of the ZPL polarize only a fraction of spins. The range of narrow linewidth OSP measurements is limited by the availability of lasers and commercial nonlinear crystals. We note that the 484~nm peak, which falls outside of our scanning range, has been previously reported to be an oxygen-related center \cite{ruan1993oxygen} and is consistent with the observed broadband OSP. Therefore, it is a promising candidate for the WAR5 ZPL.

Previous work has hypothesized the ZPL of WAR5 to be at 543~nm \cite{cann2009magnetic}. In PL, we indeed observe a strong emission line at 543~nm, not only in the WAR5 sample, but also in samples grown in a similar oxygen chemistry and samples implanted with oxygen [Fig.~\ref{fig:pl}(c)]. Hence, we confirm the origin of the 543~nm peak to be an oxygen-related defect. We further use sum frequency generation to produce a tunable narrowband laser source around 543~nm, and we observe no resonant feature in OSP for WAR5 [Fig.~\ref{fig:opticalesr}(b)]. Combining this observation with the broadband OSP response [Fig.~\ref{fig3}(b)], where the OSP falls off at wavelengths higher than $\sim$500nm, we eliminate the possibility of  543~nm being the ZPL of WAR5. Instead, using DFT calculations, we tentatively assign this optical line to the positively charged state of oxygen vacancy, OV$^+$ (Appendix \ref{appendix:ovplus}).

Finally, the ZPL emission for \ov{} may be weak because of competition with non-radiative pathways. This may be masked by other brighter emission lines in our sample, and the actual ZPL may be beyond the list of PL peaks observed in this work. The WAR5 spin signature is not observed in ESR in other oxygen-implanted and oxygen-grown samples, where the number of polarizable spins may fall below the sensitivity limit. Successful formation of the WAR5 defect in diamonds with no PL background would facilitate a clear observation of its emission spectrum and other optical properties.

\subsection{Electronic structure of \ov{}}
\begin{figure*}[httbp]
\includegraphics[width=1\linewidth]{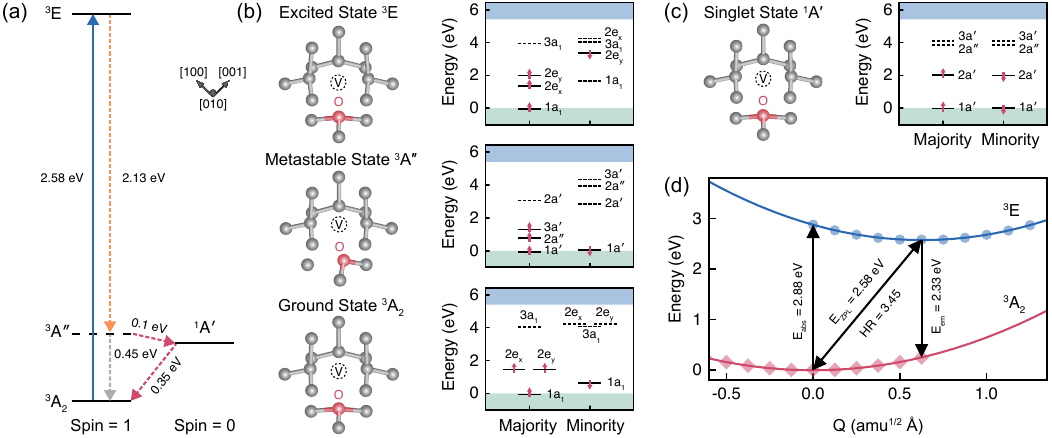} 
    \caption{Atomic and electronic structure of \ov{}. (a) Energy levels of \ov{}, including both triplet (spin = 1) and singlet (spin = 0) states. (b) Atomic configurations and Kohn-Sham levels of the triplet states, including the $^3A_2$ ground state ($C_{3v}$ symmetry), $^3A^{\prime\prime}$ metastable state ($C_s$ symmetry), and $^3E$ excited state. (c) Atomic configurations and Kohn-Sham levels of the $^1A^\prime$ singlet state. Carbon atoms are denoted in gray and oxygen in red. Green and blue shaded bands indicate the valence band and conduction band, respectively. (d) Configuration coordinate diagram of the $^3A_2$ and the $^3E$ states. The horizontal axis represents the generalized coordinate $Q$, and the difference in $Q$ indicates the difference between the ground state and excited state geometries. The absorption, emission, and zero-phonon line energies are denoted as $E_\text{abs}$, $E_\text{em}$, and $E_\text{ZPL}$, respectively, and the Huang-Rhys factor as HR.}
    \label{fig4}
\end{figure*}

To develop a model for the OSP mechanism, first-principles calculations for \ov{} are performed with the framework of the generalized Kohn-Sham (KS) theory \cite{fuchs2007quasiparticle} as implemented in the VASP code \cite{kresse1996efficient} (Appendix \ref{appendix:dft}). \ov{} has a spin-triplet ground state ($^3A_2$) with $C_{3v}$ symmetry, is isoelectronic with NV$^-$, and exhibits a similar electronic structure. The relevant energy levels of \ov{} are shown in Fig.~\ref{fig4}(a), along with the occupation of defect-induced KS states in the band gap [Fig.~\ref{fig4}(b, c)]. Starting from the $^3A_2$ ground state, the lowest spin-conserving dipole-allowed transition involves moving an electron from $1a_1$ to $2e$ in the spin-minority channel, yielding a $^3E$ excited state. The $^3A_2\rightarrow$ $^3E$ transition is allowed with a transition dipole moment $\mu=0.25\ e\cdot$\AA{}, and has a ZPL energy of 2.58~eV (480.6~nm) and Huang-Rhys (HR) factor of 3.45 [Fig.~\ref{fig4}(d)]. The calculated vertical excitation energy [$E_{abs}$ in Fig.~\ref{fig4}(d)] is 2.88~eV (430.5~nm). Taking this as the center of the phonon sideband, we expect the absorption spectrum to span the range of 2.58 to 3.18~eV (480.6 to 389.9~nm). The calculated ZPL and absorption sideband are in good agreement with the experimental data [Fig.~\ref{fig3}(b)]. Proximity to the absorption peak also explains why the spin polarization with 455~nm (2.72 eV) is more effective than 405~nm (3.06 eV) [Fig.~\ref{fig3}(a)].

In principle, ZPL emission could be observable in \ov{}; we calculate a radiative rate of $3\times10^6$~s$^{-1}$. In addition, the $^3E\rightarrow$ $^3A_2$ transition competes with the transition to a metastable triplet state $^3A^{\prime\prime}$ with reduced $C_s$ symmetry, with one elongated and two shortened C-O bonds [Fig.~\ref{fig4}(b)]. We call $^3A^{\prime\prime}$ metastable because it has a similar electron occupation as $^3A_2$, and it is locally stable. However, the different symmetry modifies the KS states and the energy of the $^3A^{\prime\prime}$ state is 0.45 eV higher than the $^3A_2$ ground state. $^3A^{\prime\prime}$ can decay to $^3A_2$ non-radiatively. Accurate calculation of rates associated with the $^3A^{\prime\prime}$ state is complicated by the unusual electronic rearrangements; however, we can assume that the $^3A^{\prime\prime}$ state will have a finite population. 

In addition to the triplet states, \ov{} also has a singlet state at an energy 0.35~eV above the $^3A_2$ ground state, as shown in Fig.~\ref{fig4}(c). The state has $C_s$ symmetry, and we label it $^1A^\prime$. The presence of this $^1A^\prime$ state allows for an intersystem crossing and hence for the experimentally observed spin polarization. The large energy difference (2.23~eV) makes a direct transition between $^3E$ and $^1A^\prime$ unlikely (we cannot exclude the existence of a singlet excited state in the energy range between $^3E$ and $^1A^\prime$, but we have not been able to stabilize such a state). However, as shown in Fig.~\ref{fig4}(a), the metastable $^3A^{\prime\prime}$ state lies at an energy 0.1~eV higher than the $^1A^\prime$ state, indicating that a spin-flip process ($^3A^{\prime\prime} \rightarrow$ $ ^1A^\prime \rightarrow\ ^3A_2$) can therefore occur, which will compete with the spin-conserving decay ($^3A^{\prime\prime}\rightarrow\ ^3A_2$). Overall, this introduces spin-selectivity and provides a unified explanation for the slow but accessible OSP observed for the WAR5 defect.

\section{Summary and Outlook}
In summary, we measure the spin properties of WAR5 across a range of temperatures, and we find that it has millisecond-scale spin lifetime at room temperature. The $T_2$ observed in this work is much lower than the 2$T_1$ limit, limited by noise arising from the high-density P1 center bath in the sample, indicating that fabricating WAR5 samples with fewer background impurities should allow for much longer coherence times. The key challenge for future work lies in identifying the ZPL of WAR5, for example, by employing a widely tunable narrow laser source for OSP measured through ESR. Knowledge of the ZPL would open up pathways for establishing optical readout in WAR5. Moreover, defects like oxygen vacancy also offer insights into design principles for new color centers in diamond with good spin properties; similar to NV, the \ov{} center contains light atoms with low spin-orbit coupling, small ionic radii that reduce lattice distortion, and an $S=1$ spin ground state with a balanced spin configuration \cite{edmonds2008electron}. In contrast, negatively-charged group IV color centers, which are spin $\frac{1}{2}$ defects, suffer from phonon-mediated orbital relaxation \cite{jahnke2015electron}, resulting in rapid spin dephasing at elevated temperatures. Future searches for long spin lifetime in diamond should focus on defects that incorporate light heteroatoms in the top rows of the periodic table.

\begin{acknowledgments}
 We gratefully acknowledge Alex Abulnaga for helpful feedback on the experiments, Gordian Fuchs for valuable suggestions on cryogenics and electronics, and Arunava Das and Alexander Hilty for help with testing the experimental setup. The experimental (S.M., A.L., J.E., S.K., Z.-H.Z., and N.P.d.L) and theoretical (Y.L., T.O.P., Y.S., and M.E.F) work was primarily supported by the Center for Molecular Quantum Transduction (CMQT), an Energy Frontier Research Center funded by the U.S. Department of Energy, Office of Science, Basic Energy Sciences under Contract No. DE-SC0021314. ESR measurements (S.M., A.L., J.E., S.K., Z.-H.Z., and N.P.d.L) and computational studies (Y.C., M.E.T, M.W., and C.G.VdW.) were also supported by the U.S. Department of Energy, Office of Science, National Quantum Information Science Research Centers, Co-design Center for Quantum Advantage (C2QA) under contract number DE-SC0012704. The research used resources of the National Energy Research Scientific Computing Center, a user facility supported by the Office of Science, U.S. Department of Energy, under Contract No. DE-AC02-05CH11231 using NERSC award BES-ERCAP0021021.\\
\end{acknowledgments}

\appendix
\renewcommand{\thefigure}{A\arabic{figure}}
\setcounter{figure}{0}
\section{Experimental Methods}
\subsection{Pulsed electron spin resonance (ESR)}
\label{appendix:esr}
\begin{figure}[b] 
    \centering
    \includegraphics[width=1\linewidth]{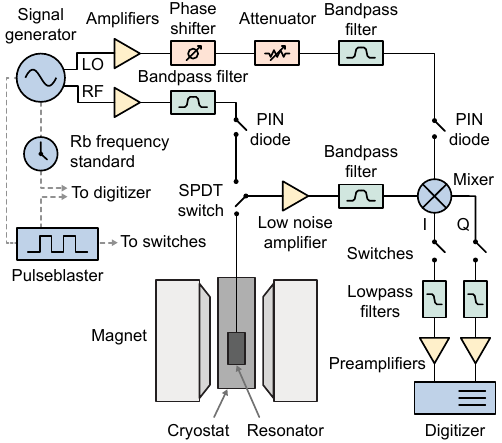}
    \caption{Schematic diagram of the home-built pulsed ESR setup. Dashed gray lines denote synchronization channels.}
    \label{fig:esrsetup}
\end{figure}

\textbf{Setup description:} Pulsed ESR measurements are performed in a home-built setup operating in the X-band ($\sim$9.7~GHz) with a sapphire dielectric ring resonator (Bruker ER 4118 X-MD-5-W1). Cryogenic temperatures are achieved with a helium flow cryostat (Oxford CF935, ColdEdge Stinger). Spins are optically polarized via a multimode fiber (Thorlabs M119L03, 400~$\mu$m core diameter) in contact with the sample, held in place with a coaxial insert (Wilmad WGS-5BL-SP). Laser excitation is pulsed with the help of a mechanical shutter (NM Laser Products LST200SLP). The magnet and its power supply (Bruker B-E 25, ER083) are controlled with a field controller and teslameter (Lakeshore F41) equipped with a Hall probe (Lakeshore FP-NS-180-TS05M).  

A schematic diagram of the microwave bridge is shown in Fig.~\ref{fig:esrsetup}. A vector signal generator (Agilent E8267D) is used for the microwave excitation and pulse modulation. A high-power amplifier (Pasternack PE15A5071, 43~dB gain) is utilized to amplify the excitation, switched on 3~$\mu$s before and after each pulse for a high extinction ratio. A bandpass filter centered at 9.625~GHz (K\&L 5C50-9625/T250-O/O, 250 MHz bandwidth) and a fast PIN diode (Pasternack PE71S5005) suppress the noise further and generate clean pulse edges. A low insertion loss single-pole double-throw (SPDT) switch (Hittite Microwave Corporation EV1HMC547ALP3) is used to switch between the excitation and detection paths during a pulse sequence. On the detection arm, the signal is amplified by a low noise amplifier (Pasternack PE15A1025) followed by a bandpass filter (Mini-circuits ZVBP-9750-S+, 9.5-10~GHz) and down-converted to DC using a mixer (Polyphase AD90120B). The local oscillator (LO) from the microwave generator goes through an amplifier (Mini-circuits ZX60-183A-S+) and a variable attenuator (ATM AF06CK-30) to get the optimum power for the mixer, a phase-shifter (ATM P1506D) for adjusting the signal between the in-phase and quadrature (I/Q) components, and finally through a bandpass filter (K\&L 5C50-9625/T250-O/O). The LO is also pulsed using a high-isolation PIN diode (RF-Lambda RFSPSTR0218G, 80~dB isolation) to suppress leakage through the mixer. After down-conversion, the signal goes through lowpass filters (Mini-circuits SLP-1.9+, DC-1.9~MHz), a preamplifier (Stanford Research Systems SR445A), and then read with a digitizer (AlazarTech ATS9626). Additional switches (Mini-circuits ZASWA-2-50DRA+) are placed on the I/Q channels to protect the preamplifier and digitizer from the reflected pulses leaking through the SPDT switch. A programmable pulse generator (PulseBlasterESR-PRO) forms the pulse sequences and synchronizes the switches with digital acquisition. A Rubidium frequency standard (Stanford Research Systems FS725) ensures coherence and phase stability across instruments for long experimental sequences.\\

\textbf{ESR alignment and fitting:} The diamond is mounted on a sapphire substrate attached to the open end of a Suprasil tube (Wilmad 722-PQ-7) with varnish (VGE-7031). This results in two angles for the sample misalignment, which can be quantified by $\theta$ and $\phi$ rotations about [001] and [1$\overline{1}$0] axes, respectively. The sample can be freely rotated along $\langle110\rangle$, and the magnetic field is aligned closest to $\langle111\rangle$ by minimizing the NV transition ($m_s=0\leftrightarrow+1$). For rotational dependence of ESR spectra, the NV and WAR5 transitions are identified by response to different wavelengths of optical polarization [Fig.~\ref{fig:esr_comparison}(a)]. Multi-Lorentzian fits are used to separate the WAR5 and NV components  [Fig.~\ref{fig1}(b)]. Then, we fit the rotational dependence of NV transitions to obtain $\theta=3.1(1)^\circ$, $\phi\approx0$, and a magnetic field offset of 0.16(9)~mT. This magnetic field calibration is verified with independent measurements of a reference sample (phosphorus donors in silicon) with a known g-factor. We note that the spectral range around $g\approx2$ is saturated with the P1 peaks, obscuring some of the NV and WAR5 transitions. Additionally, transitions with large deviation from $\langle111\rangle$ have weaker optical polarization and more overlap between WAR5 and NV, so all the WAR5 peaks cannot be distinguished. With the extracted data, the WAR5 transitions are fitted to obtain ZFS and g-factor. A weighted least squares method is used, so that spectra closer to $\langle111\rangle$, where there are more datapoints and less uncertainty, gain higher weightage. All WAR5 measurements are thereafter performed on the $m_s=0\leftrightarrow+1$ transition with the magnetic field aligned to $\langle111\rangle$, unless otherwise noted.\\

\begin{figure}[t]
    \includegraphics[width=1\linewidth]{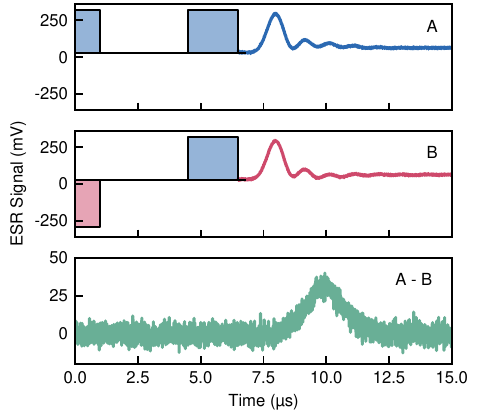} 
    \caption{Pulse sequences for phase cycling overlaid on the raw ESR signal from WAR5, recorded on a digitizer. Subtracting the traces in the top two panels (A and B) yields a clean echo signal with the cavity ring-down canceled out. Red indicates $\pi_y/2$ and blue indicates $\pi_x/2$ or $\pi_x$ pulses. Pulsewidths and amplitudes are scaled arbitrarily for illustration.}
    \label{fig:phasecycling}
\end{figure}

\textbf{Phase cycling:} High quality factors ($Q_r$) of the resonator are favorable for maximum sensitivity of narrow transitions in ESR. However, the cavity ring-down time over which the electromagnetic energy stored in the resonator cavity decays after a pulse sets a lower limit on the interpulse spacing times ($\tau$) of the echo sequences. Under critical coupling ($Q_r\sim5\times10^4$ at 4~K), ring-down effects last upto $\sim$15~$\mu$s. This can be reduced to some extent by over-coupling the resonator, at the cost of lower $Q_r$. To further shorten $\tau$, we utilize phase cycling techniques \cite{utsuzawa2022ringing}. As shown in Fig.~\ref{fig:phasecycling}, two pulse sequences are applied consecutively, with the phase of the $\pi/2$ pulse flipped. Subtracting the two traces results in a clear echo signal with ring-down distortions canceled out. A combination of over-coupling and phase cycling at 4~K allows us to achieve a higher number of pulses in dynamical decoupling, requiring short $\tau$. As the $Q_r$ is lower at room temperature, we use phase cycling with a critically coupled resonator to maximize sensitivity and measure short timescales.\\

\subsection{Nonlinear frequency conversion and supercontinuum laser}
\label{appendix:sfg}
\begin{figure}[b]
    \includegraphics[width=1\linewidth]{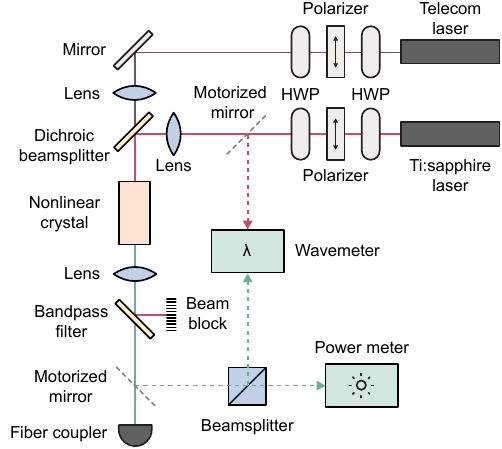} 
    \caption{Schematic diagram of the sum frequency generation setup. Two tunable lasers in near-infrared (Ti:sapphire) and telecom wavelengths are focused into a nonlinear crystal to generate tunable, narrow-linewidth optical output. For second harmonic generation, the telecom path is unused, and the dichroic beamsplitter is replaced with a mirror.}
    \label{fig:sfg}
\end{figure}

Tunable lasers around 543~nm and 491~nm were created using SFG and SHG, respectively (Fig.~\ref{fig:sfg}). For SFG, a continuous-wave Ti:sapphire laser (MSquared SolsTiS) around 836~nm and a telecom laser (Santec TSL770) with an erbium-doped fiber amplifier (OPC Photonics EYDFA-C-HP-BA-30-SM-B2) at 1550~nm are used as pump, each delivering $\sim$1~W of optical power. The beams are combined with a dichroic beamsplitter (Semrock FF925-Di01-25x36) and focused onto a 4~cm-long MgO-doped periodically poled lithium niobate (PPLN) crystal (Covesion MSFG1-0.5040). Achromatic doublet lenses are chosen according to the optimal Rayleigh range and spot size \cite{boyd1968parametric}. Polarizers and half-waveplates are used to align the polarization axes of the beams with the dipole moment of the crystal for phase matching. A wavemeter (Bristol 721) monitors both the input and the output wavelengths. The crystal is heated to an optimal temperature with an oven (Covesion PV40) and temperature controller (Covesion OC3) to attain a stable power at each wavelength. For SHG, a similar setup with only one input laser around 982 nm ($\sim$0.5~W) passing through a 2~cm-long SHG crystal (Covesion MSHG976-0.5-20) is used. Finally, the output passes through a bandpass filter (Semrock FF01-545/55-25) and is coupled to a multimode fiber fed into the cryostat. The maximum output powers for SFG and SHG are 85~mW and 12~mW, respectively, limited by the pump powers. A conversion efficiency of $\sim2.1\%$~W$^{-1}$cm$^{-1}$ is obtained for SFG, and $\sim1.2\%$~cm$^{-1}$ for SHG, consistent with manufacturer specifications. 

To probe OSP across a broad wavelength range [Fig.~\ref{fig3}(b)], ESR measurements are performed with a supercontinuum white laser source (NKT Photonics SuperK FINAIUM FIR-20) combined with tunable bandpass filters (Semrock VersaChrome TBP01-501/15-25x36, TBP01-561/14-25x36, TBP01-628/14-25x36) in different wavelength ranges. The filters are tuned with a motorized rotation stage (Thorlabs ELL18K), and central wavelengths are calibrated with a spectrometer (Princeton Instruments SpectraPro HRS-300, PIXIS 100 camera). The supercontinuum laser power is actively adjusted to maintain constant output power after the filter as the wavelength is varied. Additional points below the wavelength range of the white laser are measured with individual laser diodes at 405~nm (H\"UBNER Photonics Cobolt 06-MLD), 425~nm (Sharp GH04I01A2G), 455~nm (Thorlabs L450G3), and 473~nm (Lasermate BML473-10F1A1), set to the same optical power.

\begin{figure}[t!]
\includegraphics[width=1\linewidth]{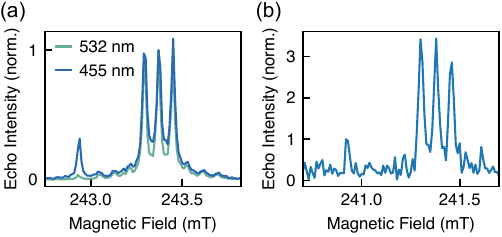} 
    \caption{(a) Comparison of ESR spectra under 532~nm and 455~nm optical polarization at 4~K, showing distinction between WAR5 and NV peaks. The spectra are normalized to the central NV peak. (b) ESR spectrum of WAR5 at room temperature with 455~nm, normalized to the WAR5 peak. The peaks at 4~K and room temperature are at different magnetic fields due to a shift in the resonator frequency.}
    \label{fig:esr_comparison}
\end{figure}
\subsection{Photoluminescence (PL) spectroscopy}
PL spectroscopy is performed in a home-built confocal microscope optimized for bulk PL detection. The sample is cooled inside a helium-flow cryostat (Janis ST-100). For excitation, we use 405~nm (H\"UBNER Photonics Cobolt 06-MLD), 473~nm (Lasermate BML473-10F1A1), and 491~nm (Edmund Optics 23-762) lasers. The excitation beam is focused onto the sample with a 10x objective (Olympus LMPLN10XIR), and PL emission is collected into a 50~$\mu\mathrm{m}$ multimode fiber routed to a spectrometer (Princeton Instruments SpectraPro HRS-300, PIXIS 100 camera). The excitation and detection paths are separated using dichroic beamsplitters (Thorlabs DMLP425R, Semrock Di03-R473-t1-25x36, Edmund Optics 86-333) or a 50:50 beamsplitter (Thorlabs BSW10R). The excitation is filtered with bandpass (Thorlabs FBH405-10, FBH473-3, FBH488-10) and the detection with longpass filters (Edmund Optics 84-742, Semrock LP02-473RU-25, LP03-532RU-25), chosen according to the excitation wavelength.\\

\section{Spin-lattice relaxation}
\subsection{Spectral diffusion of spins}
\label{appendix:T1}
\begin{figure}[t]
\includegraphics[width=1\linewidth]{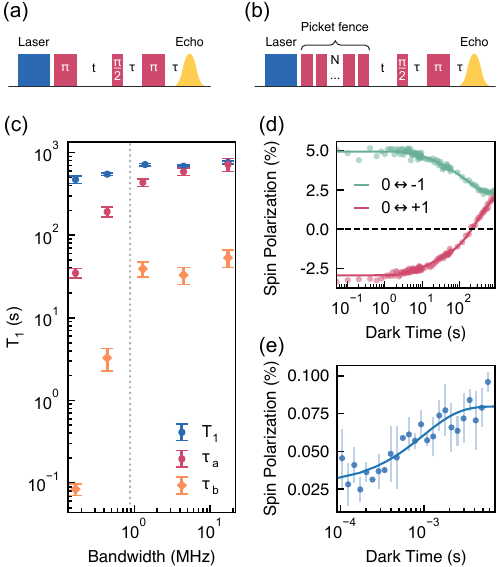} 
    \caption{Pulse sequences for $T_1$ measurement: (a) inversion recovery and (b) saturation recovery. (c) Dependence of spin relaxation time measured via saturation recovery ($T_1$) and the biexponential timescales observable in inversion recovery ($\tau_a, \tau_b$) on the bandwidth of microwave excitation. The gray dotted line indicates the ESR linewidth of WAR5. The measurements have been performed at 10~K with 455~nm optical polarization. (d) Inversion recovery on WAR5 at 10~K with 405~nm optical polarization, fit to biexponential curves. The green and red curves indicate the relaxation processes of $0\leftrightarrow-1$ and $0\leftrightarrow+1$ transitions, respectively. (e) Saturation recovery on WAR5 at room temperature. The data is fit to a single exponential with $T_1$ = 0.97(27)~ms. Error bars are standard errors from a moving average of 2 data points.}
    \label{fig:T1spectral}
\end{figure}

Conventionally, spin relaxation times are measured in pulsed ESR via inversion recovery, where a $\pi$-pulse flips the spin polarization and the relaxation from $m_s=+1$ to $m_s=0$ is measured [Fig.~\ref{fig:T1spectral}(a)]. However, we observe a biexponential curve in inversion recovery [Fig.~\ref{fig:T1spectral}(d)], which is attributed to spectral diffusion \cite{schweiger2001principles}. For excitation bandwidths lower than the ESR linewidth, only a fraction of spins are excited and measured. Spin-spin relaxation of spins beyond the measurement bandwidth changes the local field of detected spins. Spin polarization can be transferred from regions of the spectrum where spins cannot be excited or detected to the remaining spectral region, and vice versa. This is effectively observed as a secondary relaxation process. Spectral diffusion timescales shorter than $T_1$ result in a biexponential recovery curve \cite{eaton2005saturation}. For dense ensembles, dipolar coupling between these two categories of spins can also transfer polarization between them via spin diffusion \cite{schweiger2001principles}. 

For experimental verification, we first confirm that the biexponential decay is not related to the charge dynamics of WAR5 by measuring inversion recovery with thermal polarization instead of an OSP pulse, and observing the same behavior. Next, we perform $T_1$ measurements as a function of bandwidth, by varying the excitation power and hence the $\pi$ pulsewidth [Fig.~\ref{fig:T1spectral}(c)]. Bandwidths are calculated from the Fourier transform of $\pi$-pulses. Both the longer and shorter timescales in inversion recovery, $\tau_a$ and $\tau_b$, are reduced as the bandwidth decreases, and do not reflect the actual spin relaxation timescale.

The saturation recovery sequence uses a picket fence of $\pi/2$ pulses to scramble the spin polarization. Then a Hahn echo sequence detects recovery to thermal equilibrium [Fig.~\ref{fig:T1spectral} (b)]. This results in spins relaxing from equal populations of both spin levels, which mitigates spectral diffusion. The interpulse spacing of the picket fence is chosen to be shorter than the spectral diffusion timescale, but $\gg T_2$. As the bandwidth is reduced, $T_1$ measured via saturation recovery does not show a significant variation [Fig.~\ref{fig:T1spectral}(c)]. Importantly, $\tau_a$ from inversion recovery converges to the saturation recovery $T_1$ at large excitation bandwidths. However, bandwidths comparable to the WAR5 spin linewidth are favorable to avoid measuring NV spins with closely spaced energies. Hence, we use saturation recovery for reporting $T_1$. We note that saturation recovery uses only half of the population difference and offers lower sensitivity compared to inversion recovery. Inversion recovery also demonstrates that the $m_s=0$ state is preferentially populated with OSP. Fig.~\ref{fig:T1spectral}(d) shows that the spin polarization is inverted for the $0\leftrightarrow+1$ measurement but not for $0\leftrightarrow-1$, where thermal polarization relaxes spins to $m_s=-1$.

\subsection{Phonon density of states}
\label{appendix:phonon}
Density functional theory (DFT) yields the phonon dispersions and ZFS tensor, which is used to evaluate the spin-phonon coupling strength of the \ov{} defect. DFT calculations are performed using VASP (version 6.2.1~\cite{hafner2008ab}) with the PAW method and PBE exchange-correlation functional~\cite{perdew1996generalized}. We substitute a single O atom at a C site, corresponding to a doping concentration of 1.587\%, thereby creating a vacancy near the O dopant, which favors a neutral charge state. Calculations use a plane-wave kinetic-energy cutoff of 500~eV and a Gaussian smearing width of 0.1~eV. Brillouin-zone sampling uses a 2 $\times$ 2 $\times$ 2 $k$-mesh on a supercell of  2 $\times$ 2 $\times$ 2, with key phonon properties carefully examined on a larger supercell of 3 $\times$ 3 $\times$ 3. Electronic and ionic self-consistency is converged to 10$^{-9}$ and 10$^{-8}$~eV, respectively. Spin polarization is included in all calculations. Phonons are computed using the density functional perturbation theory~\cite{baroni2001phonons} with PHONOPY~\cite{togo2023first}. The ZFS tensor is obtained directly from VASP using its built-in LDMATRIX/DDIAG output from spin-polarized calculations.

\begin{figure}[t!]
\includegraphics[width=1\linewidth]{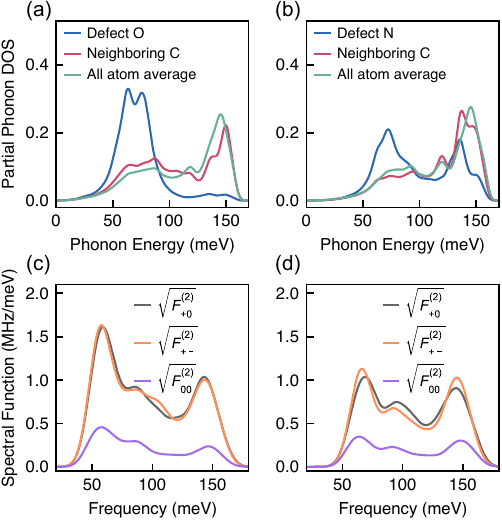} 
  \caption{Atom projected phonon density of states (DOS) for (a) \ov{}–diamond and (b) NV$^-$–diamond. Vacancy localized clusters are observed near $\sim$70~meV and $\sim$145~meV for both systems. The bulk optical modes are also centered around $\sim$145 meV, hybridizing heavily with the higher frequency localized clusters. Spectral function of coupling-weighted DOS via second-order interaction for (c) \ov{} and (d) NV$^-$, broadened with a Gaussian smearing of 7.5 meV. The coupling matrix elements shift peak locations for \ov{} to 58.8 and 143.0~meV, respectively.}
  \label{fig:pdos}
\end{figure}

The calculated partial phonon density of states (PhDOS) exhibits two prominent phonon groups centered at 69.2~meV, primarily arising from oxygen vibrations, and around 149.8~meV, dominated by carbon atoms [Fig.~\ref{fig:pdos}(a)]. In contrast, the NV center in diamond shows two quasilocalized phonon modes at approximately 68.2 and 167~meV \cite{cambria2023temperature} associated with nitrogen [Fig.~\ref{fig:pdos}(b)].

To support the result obtained from the PhDOS, we calculate the spin-phonon coupling constant using \cite{briganti2021complete} $\hat{V}_{\alpha} = \sum_{lm} (\delta B^l_m / \delta Q_{\alpha}) \hat{O}^l_m$, where $B^l_m$ denotes the crystal-field parameters, which are directly related to ZFS parameters, $D$ and $E$, via $D = 3 B_2^0$ and  $E = B_2^2$. Here, $Q_\alpha$ represents the displacement coordinate of the phonon mode $\alpha$ at the $Q$ ($\Gamma$) point. $\hat{O}^l_m$ are the Stevens operators, such that $\hat{H}_{CF} = \sum_{l}\sum_{m}B^l_m\hat{O}^l_m$. These can be expressed in terms of spin operators,
\begin{equation}
    \hat{O}_2^0 = 3 \left( \hat{S}_z^2 - \frac{\hat{\boldsymbol{S}}^2}{3} \right)
    \quad \text{and} \quad 
    \hat{O}_2^2 = \frac{(\hat{S}_+^2 + \hat{S}_-^2)}{2},
\end{equation}
where $\hat{\boldsymbol{S}}^2 = \hat{S}_x^2 + \hat{S}_y^2 + \hat{S}_z^2$ and $\hat{S}_\pm = \hat{S}_x \pm i\hat{S}_y$. The calculated spin-phonon coupling constants are used to evaluate the spectral functions, which are directly related to the transition rates, as discussed below. The temperature-dependent spin-lattice relaxation rate associated with Raman transitions, arising from first-order interactions applied to second-order in perturbation theory, can be computed using Fermi's golden rule~\cite{cambria2023temperature}, 

\begin{align}
\label{first-order_spin-phonon_rate}
W^{(1)}_{2(a,b)}(T)&\\
= \frac{2\pi}{\hbar}
\sum_{\alpha,\alpha',c}
\Bigg\{&{n}_\alpha \big({n}_{\alpha'} + 1\big)
\left|
\frac{
\langle a | V_\alpha | c \rangle
\langle c | V_{\alpha'} | b \rangle
}{
E_c - E_a - \hbar \omega_\alpha
}
\right|^2
\delta(\Delta E_{-})
\nonumber \\
+ &{n}_{\alpha'} \big({n}_\alpha + 1\big)
\left|
\frac{
\langle a | V_\alpha | c \rangle
\langle c | V_{\alpha'} | b \rangle
}{
E_c - E_a + \hbar \omega_\alpha
}
\right|^2
\delta(\Delta E_{+})
\Bigg\}\nonumber
\end{align}
where $\langle a|V_{\alpha}|c \rangle$ and $\langle c|V_{\alpha}|b \rangle$ indicate the spin-phonon coupling matrix elements for the transitions from state $a$ to $c$ and from $c$ to $b$.  $\Delta E_{\pm} = E_{b} \pm (\hbar \omega_\alpha - \hbar \omega_{\alpha^{'}}) - E_{a}$ is the energy difference between the final ($b)$ and initial ($a$) states and ${n}_\alpha=\left(e^{\hbar\omega_\alpha/k_B T}-1\right)^{-1}$ is the Bose-Einstein occupation. Eq.~\ref{first-order_spin-phonon_rate} can be further expressed in the form of spectral function $F(\hbar\omega)$:

\begin{align*}
W^{(1)}_{2(a, b)}(T)
= \frac{4\pi}{\hbar} \sum_{c} &\int_{0}^{\infty} d(\hbar \omega)\,
{n}(\omega)\,[{n}(\omega)+1]\,
\\&\times\frac{F^{(1)}_{a, c}(\hbar \omega)\,
F^{(1)}_{c, b}(\hbar \omega)}
{(\hbar \omega)^2}
\end{align*}
where $F^{(1)}_{a, b}(\hbar \omega)$ is the first-order spectral function obtained by convolving the first-order coupling coefficients. Here, we use a Gaussian smearing ($\sigma$) of 7.5~meV.
\begin{align}
F^{(1)}_{a, b}(\hbar \omega)
=&\nonumber \int d(\hbar \omega') \sum_{\alpha}
\left| \langle a|V_\alpha|b\rangle \right|^2
\delta(\hbar \omega' - \hbar \omega_\alpha)
\\&\times
\frac{1}{\sigma \sqrt{2\pi}}
\exp\!\left(
-\frac{(\hbar \omega' - \hbar \omega)^2}{2\sigma^2}
\right)
\label{spectral_function_first}
\end{align}

Similarly, the spin-lattice relaxation rate due to the second-order interactions  applied to first-order in perturbation theory can be expressed as,
\begin{equation}
W^{(2)}_{1(a, b)}(T) = \frac{4\pi}{\hbar} \int_0^\infty d(\hbar\omega) {n}(\omega)({n}(\omega) + 1) F^{(2)}_{a b}(\hbar\omega, \hbar\omega)
\end{equation}
where $\sqrt{F^{(2)}_{a b}(\hbar\omega, \hbar\omega)}$ is the second-order spectral function by convolving the
second-order coupling coefficients:

\begin{align}
\sqrt{F^{(2)}_{a b}(\hbar\omega, \hbar\omega)} = &\nonumber\int d(\hbar\omega') \sum_\alpha \left| \langle a|V_{\alpha\alpha}|b\rangle \right| \delta(\hbar\omega' - \hbar\omega_\alpha) \\
&\times \frac{1}{\sigma \sqrt{2\pi}} \exp \left( -\frac{(\hbar\omega' - \hbar\omega)^2}{2\sigma^2} \right)
\label{spectral_function_second}
\end{align}

We calculate the first and second-order coupling matrix elements and use them to compute the corresponding spectral functions using Eqs.~\ref{spectral_function_first} and \ref{spectral_function_second}. For both OV$^0$ and NV$^-$, we find the second-order interaction to be the dominating term, in agreement with Ref.~\citealp{cambria2023temperature}. The resulting spectral functions exhibit similar features for both systems. The first peak in the second-order spectral function of OV$^0$ appears around 58.8~meV, coinciding with the PhDOS peak near 69.2~meV, primarily contributed by oxygen vibrations. A second prominent peak is observed near 143.0~meV, corresponding to a PhDOS peak near 149.8~meV with a relatively larger contribution from neighboring C atoms than from O atoms. These energies (58.8~meV and 143.0~meV) are in good agreement with the experimental values of Orbach activation energies, 54.3(8.7)~meV and 137.8(10.9)~meV, extracted from the temperature dependence of $T_1$.

\section{Spin counting}
\label{appendix:spincounting}
We perform spin counting with pulsed ESR to quantify the concentration of spins in the WAR5 sample. An isotopically enriched $^{28}$Si crystal doped with phosphorus is used as a reference, with $8\times10^{10}$ spins previously calibrated with a known standard. The free evolution time in the Hahn echo sequence is set to be much shorter than $T_2$ so that the echo intensity is not diminished by decoherence. Since WAR5 and NV have extremely long $T_1$, the saturated echo intensity at thermal equilibrium is fitted from saturation recovery measurements to minimize the uncertainty. To account for the different inhomogeneous spin linewidths of different defects, the saturated echo intensity is scaled by the integrated spectral amplitude to yield the effective intensity, $I$. This incorporates the different numbers of hyperfine lines for each defect as well. The number of spins, $N_\text{spins}$, is given by \cite{weil2007electron}

\begin{equation}
N_\text{spins} \propto \frac{I\ n_\text{sites}}{\rho \sqrt{Q_r} \sqrt{S(S+1)-(S-1)S}}
\label{eq:spincounting}
\end{equation}
where $n_{\text{sites}}$ is the number of inequivalent orientations of the defect, $\rho$ is the polarization fraction at thermal equilibrium, $Q_r$ is the resonator quality factor with the sample inserted, and $S$ denotes the spin. $\rho$ is dependent on the temperature, resonator frequency, and spin levels. It is defined as $\frac{p_0-p_1}{p_0+p_1+p_{-1}}$ for $S=1$ and $\frac{p_{-\sfrac{1}{2}}-p_{\sfrac{1}{2}}}{p_{-\sfrac{1}{2}}+p_{\sfrac{1}{2}}}$ for $S=\sfrac{1}{2}$, where $p_{m_s}$ is the population of the $m_s$ spin state from Boltzmann distribution. We obtain [WAR5] $\approx4\times10^{15}$~cm$^{-3}$, [NV] $\approx1.7\times10^{16}$~cm$^{-3}$, and [P1] $\approx5\times10^{17}$~cm$^{-3}$. From natural abundance, [$^{13}$C] $\approx1.1\%=1.9\times10^{21}$~cm$^{-3}$. The single-shot detection limit for the ESR setup is characterized to be $\sim3\times10^9$ spins.

\section{Decoherence}
\subsection{Cluster-correlation expansion}
\label{appendix:cce}
Now, we calculate the stretching factor of WAR5 under the observed spin bath using the generalized cluster-correlation expansion (gCCE) method \cite{onizhuk2021probing}. This is implemented with the help of the PyCCE package in Python \cite{onizhuk2021pycce}. A diamond crystal unit cell is expanded to a cubic supercell with an edge length of 3000~\AA{}. The central spin is an oxygen vacancy ($^{16}$O) with the experimental values of the ZFS and magnetic field. The maximum number of spins interacting within a cluster is dictated by the CCE order. We implement gCCE3, as the 2nd order is not a good approximation for strongly coupled electron spin baths. Since the stretching factor calculated with gCEE3 has been shown to be independent of the P1 spin concentration \cite{ghassemizadeh2024coherence} in our range, we simulate a lower P1 concentration of $5\times10^{16}$  cm$^{-3}$ for the sake of computational cost. Natural abundance of $^{13}$C is assumed. To ensure convergence, we choose $r_\text{dipole}=300$~\AA{} and $r_\text{bath}=600$~\AA{}  \cite{ghassemizadeh2024coherence}, which specify the maximum distance between two bath spins in a cluster and the cutoff radius from the central spin, respectively. Contribution from bath spins outside a cluster is taken into account via Monte Carlo sampling over 100 random initial bath states. 500 spatially different configurations of bath spins are simulated, out of which 100 configurations are ensemble-averaged over 1000 random combinations. The mean decays are fit to a stretched exponential to generate the histogram in Fig.~\ref{fig:T2noise}(c). The calculated stretching factor is in excellent agreement with our experimental observations.

\subsection{Noise spectroscopy}
\label{appendix:noise}
\begin{figure}[b]
\includegraphics[width=1\linewidth]{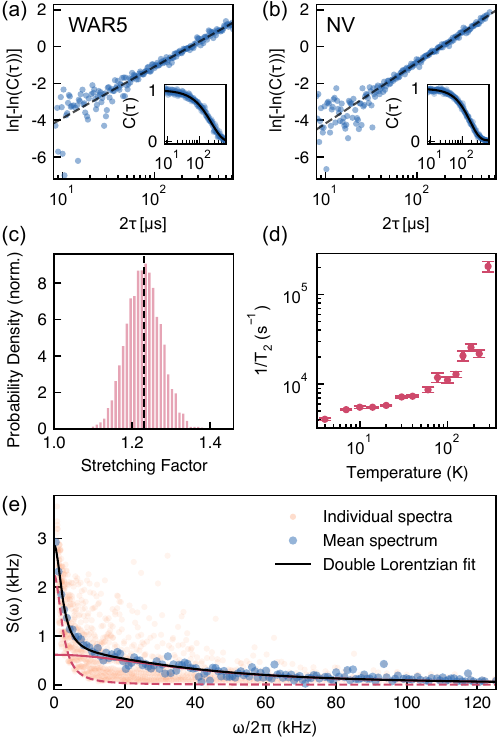} 
    \caption{Hahn echo $T_2$ at 4~K for (a) WAR5 and (b) NV centers. The log-log vs log scale plots show the stretching factor as a slope. Insets show the coherence, $C(\tau)$, on a linear scale. Black dashed and solid lines are fits to a stretched exponential, $e^{-(t/T_2)^\beta}$, where $t=2\tau$. WAR5: $T_2$ = 246(7)~$\mu$s, $\beta$ = 1.22(4). NV: $T_2$ = 183(2)~$\mu$s, $\beta$ = 1.44(3). (c) gCCE simulation of \ov{} yielding a distribution of stretching factors, with a mean value of $\beta\approx1.23$. (d) Hahn echo $T_2$ of WAR5 as a function of temperature. (e) Noise spectrum of WAR5 from spectral decomposition. Orange and blue dots indicate the individual traces computed for every $N_\pi$ and the mean spectrum, respectively. The black line is a fit to a double Lorentzian model (Eq. \ref{eq:lorentzian}). Solid and dashed red lines are the individual Lorentzian spectra.}
    \label{fig:T2noise}
\end{figure}

Spectral decomposition is a useful technique to study spin-bath dynamics. The CPMG sequence can be used as a filter function to probe different frequencies in the noise spectrum. The coherence decay of a spin is modeled by $C(t) = e^{-\chi(t)}$, where
\begin{equation}
\label{eq:chi}
    \chi(t)=\frac{1}{\pi}\int_0^\infty d\omega S(\omega) \frac{F_N(\omega t)}{\omega^2}
\end{equation}
for noise spectrum $S(\omega)$. Approximating the CPMG filter function $F_N(\omega t)$ as sharply peaked,
\begin{equation}
\label{eq:chiapprox}
    \chi(t)\approx\frac{tS(\omega)}{\pi}
\end{equation}
As $T_{2,\mathrm{CPMG}}$ is proportional to $N_\pi^\gamma$, $N_\pi$ being the number of $\pi$-pulses in the CPMG sequence, 
\begin{equation}
\label{eq:chiT2}
    \chi(t)=(t/T_{2,\mathrm{CPMG}})^\beta\propto(t/N_\pi^\gamma)^\beta
\end{equation}
for stretching factor $\beta$. Consider the noise spectrum to follow a power law, $S(\omega)\propto \omega^{-\alpha}$. Comparing powers of the independent variable $N_\pi$ in Eq.~\ref{eq:chiapprox} and \ref{eq:chiT2}, we get,
\begin{align}
    &\frac{t^\beta}{N_\pi^{\gamma\beta}} \propto \frac{t}{(N_\pi/t)^\alpha}\implies\gamma=\frac{\alpha}{\beta}
\label{eq:scaling}
\end{align}
The noise spectrum computed from Eq.~\ref{eq:chiapprox} is plotted in Fig. \ref{fig:T2noise}(e). Noise exponent, $\alpha\approx0.64$ from a linear fit of $\ln[S(\omega)]$ vs $\ln(\omega)$. Stretching factor for WAR5 is $\beta\approx1.22$ [Fig.~\ref{fig:T2noise}(a)]. Thus, from Eq.~\ref{eq:scaling}, we get $\gamma\approx0.52$, i.e. $T_{2,\mathrm{CPMG}}\propto\sqrt{N_\pi}$.

While the power law is a simplified approximation for the noise that explains the CPMG scaling, we find the best fit to the WAR5 noise spectrum to be a double Lorentzian model:
\begin{equation}
S(\omega)=\sum_{i=1,2}\frac{\Delta_i^2 \tau_{c_i}}{\pi \left[ 1+(\omega \tau_{c_i})^2 \right]}
\label{eq:lorentzian}
\end{equation} 
where $\Delta_i$ is the coupling strength between the central spin and the spin bath, $\tau_{c_i}$ is the correlation time of the spin bath related to their characteristic flip-flop time, and the subscript $i$ denotes different spin species. $\Delta_i$ is given by the average dipolar interaction energy between the bath spins and WAR5 and scales as the spin density, whereas $\tau_{c_i}$ is given by the inverse of the dipolar interaction energy between neighbouring bath spins and is inversely proportional to their density. For the WAR5 noise spectrum, we obtain $\Delta_1=8.9 (1)$~kHz, $\tau_{c,1}=24.4(1.4)$~\us, $\Delta_2=4.2(1)$~kHz, and $\tau_{c,2}=405.6(24.9)$~\us. NV measurements in the WAR5 sample also yield a double Lorentzian noise spectrum with similar values. We can rule out the possibility of NV centers contributing to the WAR5 decoherence, as their density is 29 times lower than P1, which is not reflected in the ratio of coupling strengths. Moreover, we note that the ratio of coupling strengths is an order of magnitude different from the ratio of correlation times. This can only be explained if the two spin baths originate from electron and nuclear spins, respectively. From spin counting (Appendix \ref{appendix:spincounting}), we know that the $^{13}$C and P1 spin densities are in the ratio $3.9\times10^3$. Gyromagnetic ratio for $^{13}$C nuclei is a factor of $2.6\times10^3$ lower compared to that of electrons. Further, electron-nuclear spin interaction due to $^{13}$C can suppress the P1 flip-flop rate by an order of magnitude as it broadens the electron spin linewidth \cite{bauch2020decoherence}. This does not affect the flip-flop rates for nuclear spins which are closer to each other and experience the same hyperfine environment. Combining these factors, we estimate $\Delta_{^{13}\text{C}}/\Delta_\text{P1}\sim1.5$ and $\tau_{c,\text{P1}}/\tau_{c,^{13}\text{C}}\sim15$ from spin densities. This is consistent with the parameters from the noise spectrum, $\Delta_{1}/\Delta_2\approx2.1$ and $\tau_{c,2}/\tau_{c,1}\approx16.6$. Hence, we attribute the noise bath to P1 centers and $^{13}$C nuclear spins.\\

\section{PL spectroscopy on WAR5}
\label{appendix:pl}
\begin{figure}[b!]
\includegraphics[width=1\linewidth]{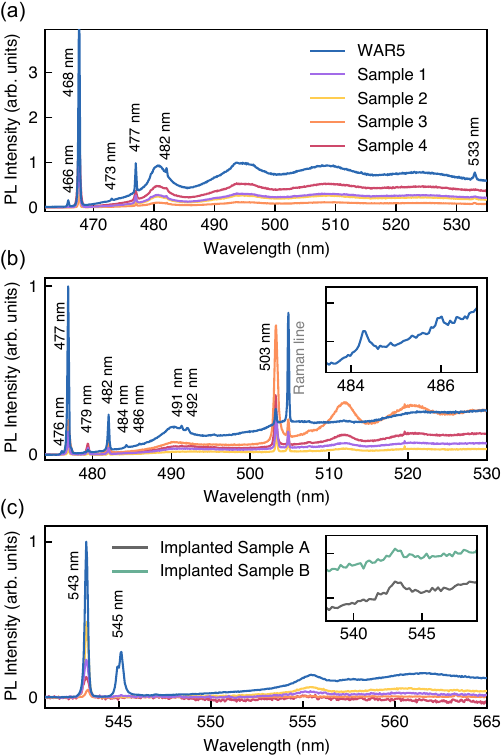} 
    \caption{PL spectra of WAR5 and other oxygen-grown samples (labeled 1 to 4) at 10~K with (a) 405~nm, (b) 473~nm, and (c) 491~nm excitation. Inset in (b) shows a magnified version of the 484~nm and 486~nm peaks that are unique to the WAR5 sample. Inset in (c) shows that the 543~nm peak is observed in oxygen-implanted samples as well. WAR5 PL is scaled down by a factor of 10 in (a) and (b) and by 50 in (c), compared to samples 1–4. The spectra are normalized to the 477~nm peak in (a) and (b), and the 543~nm peak in (c).}
    \label{fig:pl}
\end{figure}

To search for the ZPL of WAR5, PL is measured on the WAR5 sample and compared to diamonds grown in oxygen similar to WAR5 (Samples 1 to 4) and oxygen-implanted samples A and B ($7\times10^{14}$ cm$^{-2}$ total fluence, 1200$^\circ$C annealed). A list of peaks with possible origin from literature is reported in Table~\ref{tab:pl}. Excitation wavelengths of 405~nm and 473~nm reveal spectra of emission lines in the blue region [Fig.~\ref{fig:pl} (a),(b)]. With 405~nm, the dominant feature is the 468~nm center, which is commonly observed in as-grown CVD diamonds and related to vacancy clusters \cite{zaitsev2021luminescence}. Some other peaks are reported in literature but not definitively assigned to a known defect. We also observe some of the unreported lines in other boron or nitrogen-doped diamonds not related to oxygen (marked with an asterisk in Table \ref{tab:pl}), including 491~nm and 492~nm. Eliminating all such lines, the peak at 484~nm is unique to WAR5 and can be considered a potential candidate for WAR5 ZPL. We note that the 484~nm peak has been reported in CVD diamond films grown in oxygen \cite{ruan1993oxygen} and in natural diamonds \cite{wang2018spectroscopic}. 

{\hyphenpenalty=5000
\begin{table}[t]
    \centering
    \begin{tabular}{lll}
    \toprule\toprule
    Wavelength (nm)  & Energy (eV)  & Prior reports \\\midrule
    465.8 & 2.662 & –\\
    467.6 & 2.652 & Vacancy cluster \cite{zaitsev2021luminescence}\\
    473.0* & 2.621 & Nickel-related defect \cite{zaitsev2013optical},\\
    &&not applicable for this work\\
    476.1* & 2.604 & –\\
    477.0* & 2.599 & Unknown origin in\\&&synthetic diamonds \cite{zaitsev2013optical}\\
    477.3* & 2.598 & –\\
    479.5* & 2.586 & –\\
    482.0* & 2.572 & Unknown origin in\\&&CVD films \cite{zaitsev2013optical,kanda2003characterization}\\
    484.3 & 2.560 & Oxygen-related defect \cite{ruan1993oxygen}\\
    486.0 & 2.551 & Irradiation defect \cite{zaitsev2013optical}\\
    491.3* & 2.524 & Irradiation defect \cite{rakhmanova2022spectroscopic, manfredotti2010luminescence}\\
    492.0* & 2.520 & Irradiation defect \cite{rakhmanova2022spectroscopic}\\
    503.2 & 2.464 & H3 center (NVN$^0$) \cite{ashfold2020nitrogen, johnson2025nitrogen}\\
    503.4 & 2.463 & 3H center (interstitial\\&&carbon) \cite{steeds19993h}\\
    533.0 & 2.326 & Nitrogen-related defect\\&&\cite{ruan1991cathodoluminescence,lobaev2020diamond}\\
    543.2 & 2.282 & Oxygen-related defect \cite{cann2009magnetic}\\
    545.1 & 2.275 & Nitrogen-related or\\&&irradiation defect \cite{zaitsev2013optical}\\
    \bottomrule\bottomrule
    \end{tabular}
    \caption{List of PL peaks observed in the WAR5 sample with possible origin from prior reports. Asterisks indicate peaks that we also observe in diamonds not related to oxygen.}
    \label{tab:pl}
\end{table}}

With 491~nm excitation [Fig.~\ref{fig:pl}(c)], all samples show emission at 543 nm as reported in Ref.~\citealp{cann2009magnetic}. The common emission of 543~nm in all oxygen samples confirms that this is an oxygen-related defect. In the WAR5 sample, we also observe emission at 545 nm, which may be a nitrogen-related defect or damage center \cite{zaitsev2013optical}. NV$^0$ (575~nm) and NV$^-$ (637~nm) are also observed in the oxygen-grown samples. None of the samples show emission from the oxygen-related ST1 center (535-570~nm) \cite{luhmann2022identification, pezzagna2024polymorphs}.

\section{Optical response of ESR}
\label{appendix:opticalesr}
\begin{figure}[httbp!]
    \includegraphics[width=1\linewidth]{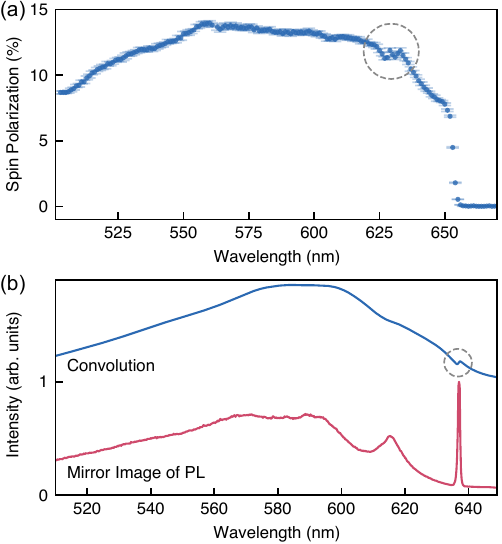} 
    \caption{(a) OSP as a function of wavelength for NV centers in the WAR5 sample. Spins are polarized for 1~s with 60~mW optical power. Error bars are standard errors from averaging 4 measurements. (b) Convolution of a 20~nm excitation band with the mirror image of PL mimics the qualitative trend of OSP observed in ESR. Dashed gray circles indicate the resonant feature around the ZPL.}
    \label{fig:nvosp}
\end{figure}

OSP with a broadband supercontinuum laser has been demonstrated for WAR5 in Fig.~\ref{fig3}. We substantiate our interpretation by measuring the OSP of NV centers with the same configuration. Due to the broad bandwidth of the filter (20~nm), the features of the absorption sideband are broadened, and a small resonant feature is visible around the ZPL [Fig.~\ref{fig:nvosp}(a)]. This is qualitatively similar to the convolution of the mirror image of PL with a 20~nm excitation band, as shown in Fig.~\ref{fig:nvosp}(b). The resonant feature is slightly shifted from the NV ZPL (637~nm) due to the inhomogeneity of the excitation spectrum. Beyond the ZPL, spin polarization drops to thermal polarization. 

\begin{figure}[t!]
    \includegraphics[width=1\linewidth]{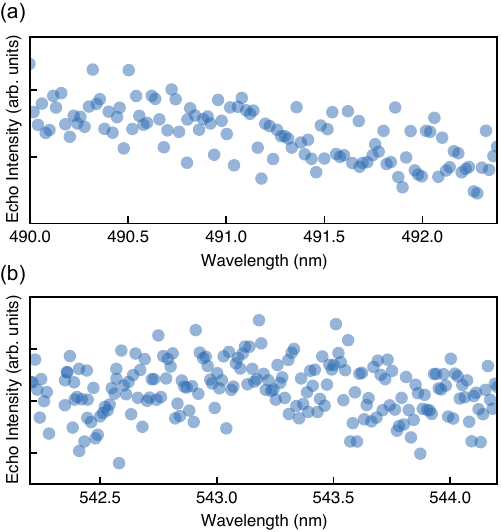} 
    \caption{Optical response of the WAR5 ESR signal around (a) 491~nm using second harmonic generation and (b) 543~nm using sum frequency generation. Optical power is stabilized at 12 mW and 70 mW, respectively. The slope observed in (a) is an artifact from resonance drift in the helium flow cryostat, verified by sweeping the wavelength in the reverse direction. }
    \label{fig:opticalesr}
\end{figure}

To further investigate the optical response of ESR, tunable narrow linewidth excitation across regions of interest is generated from sum frequency and second harmonic processes (Appendix \ref{appendix:sfg}). The output linewidth is dictated by the pump lasers ($<100$~kHz), and we measure an upper bound of 2~GHz, limited by the resolution of the spectrum analyzer (Bristol 721). The echo intensity measured on the $m_s=0 \leftrightarrow +1$ transition is recorded as a function of the excitation wavelength. Fig.~\ref{fig:sfg} demonstrates no response observed around 491 nm, 492~nm, or 543 nm. Additional measurements spanning 489–494~nm and 536–544.2~nm in parts also do not show any resonant feature. 

\section{Theory}
\subsection{Methods}
\label{appendix:dft}
First-principles calculations are performed with the framework of the generalized Kohn-Sham (KS) theory \cite{fuchs2007quasiparticle}, utilizing projector augmented-wave (PAW) potentials \cite{blochl1994projector} as implemented in the VASP code \cite{kresse1996efficient}. The hybrid functional of Heyd, Scuseria, and Ernzerhof (HSE06) \cite{heyd12452} is used, with the default mixing parameter $\alpha=0.25$ and screening parameter $\omega = 0.2$~\AA{}$^{-1}$. For the primitive cell of bulk diamond, a $\Gamma$-centered $8\times8\times8$ $k$-point mesh is used, yielding an optimized lattice constant of 3.54~\AA{}. The calculated band gap of 5.41~eV agrees well with the experimental value of 5.47~eV \cite{yamanaka1994influence}. Defect calculations are performed in a $3\times3\times3$ supercell (216 atoms) with $\Gamma$-point-only sampling. Spin polarization is included in all calculations. Structural relaxations are performed using HSE06 until the forces fall below $10^{-2}$~eV/\AA{}, with an electronic self-consistency threshold of $10^{-4}$~eV. A plane-wave energy cutoff of 400~eV is employed for all defect calculations. Increasing the energy cutoff to 520~eV or the supercell size to $4\times4\times4$ (512 atoms) results in changes in ZPL energies of less than 0.05~eV, validating the $3\times3\times3$ supercell (216 atoms) and 400~eV cutoff. 

To study internal transitions, we employ the $\Delta$SCF methodology \cite{jones1989density}. In this approach, excitation energies are obtained as total energy differences between two calculations with constrained occupations, each including full atomic relaxation. We use configuration coordinate diagrams to examine the coupling of these electronic transitions to lattice vibrations \cite{dreyer2018first}, and to compute the Huang-Rhys (HR) factors within the one-dimensional approximation \cite{alkauskas2012first}, as implemented in the NONRAD code \cite{turiansky2021nonrad}.

\subsection{OV$^+$ as the origin of the 543~nm PL center}
\label{appendix:ovplus}

\begin{figure}[b!]
\includegraphics[width=1\linewidth]{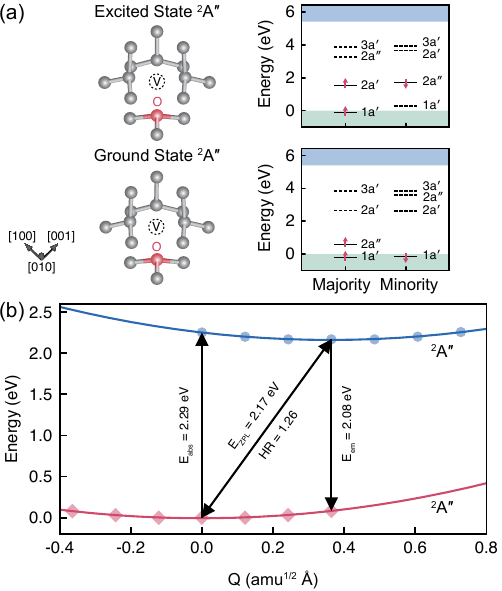} 
    \caption{Atomic and electronic structure of OV$^+$. (a) Atomic configuration and Kohn-Sham states of the doublet ground state $^2A^{\prime\prime}$ ($C_s$ symmetry, lower) and doublet excited state $^2A^{\prime\prime}$ (upper). Carbon atoms are denoted in gray and oxygen in red. Green and blue shaded bands indicate the valence band and conduction band, respectively. (b) Configuration coordinate diagram of the $^2A^{\prime\prime}$ ground state and the $^2A^{\prime\prime}$ excited state.}
    \label{fig:ovplus}
\end{figure}

From experimental evidence, we have rejected 543~nm as the ZPL of WAR5 or OV$^0$. Here, we suggest that a different charge state of the oxygen vacancy center, namely OV$^+$, can be the origin of the 543~nm (2.28~eV) PL line observed in oxygen-containing samples. Our calculations for OV$^+$ show that the $^2A^{\prime\prime} \rightarrow\ ^2A^{\prime\prime}$ transition is dipole-allowed, with a ZPL energy of 2.17 eV [Fig.~\ref{fig:ovplus}(b)], within the computational error bar of the 2.28~eV emission. Fig.~~\ref{fig:ovplus}(a) shows the KS states of the $^2A^{\prime\prime}$ ground state and the $^2A^{\prime\prime}$ excited state. The excited $^2A^{\prime\prime}$ configuration is formed by promoting an electron from $1a^\prime$ to $2a^{\prime\prime}$ in the spin-minority channel. We note that this excitation is distinct from the one reported in a previous study \cite{thiering2016characterization}, which promoted an electron from $1a^\prime$ to $2a^\prime$ in the minority channel. For that transition, we obtain a higher ZPL energy of 2.20~eV and a larger HR factor of 2.76. Our study identifies the $1a^\prime\rightarrow2a^{\prime\prime}$ excitation as the lowest-energy dipole-allowed transition of OV$^+$. Its HR factor of 1.26 indicates weak electron-phonon coupling, consistent with the narrow emission in PL [Fig.~\ref{fig:pl}(c)]. 

Experimentally, the 543~nm line is strongly enhanced in the WAR5 sample, suggesting that they share a common microscopic origin. In equilibrium, one expects a single charge state of a particular defect to dominate. However, illumination (which is necessary to observe PL) can generate other charge states. In a wide-bandgap material such as diamond, such charge states can persist for extended periods of time, as is well established for the charge states of the NV
center \cite{pederson2024rapid}. The positive charge state, OV$^+$, is thus quite likely to be present in samples that also contain \ov{}, consistent with the observation of the brightest 543~nm peak in PL in the WAR5 sample, compared to other samples.

\bibliographystyle{apsrev4-1}
\bibliography{main_WAR5}
\end{document}